\documentclass[12pt]{article}
\usepackage{amsmath,amssymb}

\begin{document}

\title{Energy and angular momentum in strong gravitating systems}

\author{Roh-Suan Tung\\
\it Center for Astrophysics\\
\it Shanghai Normal University\\
\it Shanghai 200234, China\\
tung@shnu.edu.cn}

\date{}

\maketitle

\begin{abstract}
A quasilocal framework of stationary and dynamical untrapped hypersurfaces
is introduced to generalize the notions of energy and angular momentum
of isolated and dynamical trapping horizons to general strong gravitating
systems.
\end{abstract}

\section{Introduction}
The notions of energy and angular momentum for weak gravitating systems
in classical general relativity are well understood in terms of the
symmetry at the asymptotically spatial infinity and the asymptotically
null infinity.
The question of how to define energy and angular momentum
for strong
gravitating systems has been raised for a while in searching for the
``quasi-local energy-momentum and angular momentum''
\cite{Szabados-Review}. The idea is to find a suitable
definition of total energy-momentum and angular-momentum, surrounded
by a spacelike two dimensional surface $S$, in four dimensional
spacetime $M$. The construction is quasi-local in the sense that it
refers only to the geometry of $S$ (intrinsic metric, first
fundamental form), the extrinsic curvatures (second fundamental
forms) and the connection 1-forms on the normal bundle (normal
fundamental forms) for its embedding in $M$.

An especially interesting example of strong gravitating systems
is black hole, which is now believed to be common in the universe.
Traditional description of black holes in terms of
event horizons is inadequate for the expected observational data.
For non-stationary spacetimes,
quasi-local notions of trapped and marginal surfaces
have now been found to be more useful within the framework of
isolated,
and dynamical trapping horizons \cite{Hayward1994, IH2000,
Ashtekar-Krishnan2002,Ashtekar-Krishnan2003,
Gourgoulhon2005, Booth-Fairhust2005,
Hayward2006}. These frameworks enable one to significantly extend
the laws of black hole mechanics to the dynamical regime with the
associated notions of energy, angular momentum and their fluxes, and
have been applied to several problems in mathematical general
relativity, numerical relativity, and quantum gravity
\cite{Ashtekar-Krishnan-Review}.
These progresses on black hole dynamics lead to a question whether
we can generalize the conservation laws for
isolated and dynamical
trapping horizons to general untrapped regions so that we can study
the change of energy, angular momentum and their fluxes for
untrapped strong gravitating systems, e.g., before the black hole
horizon was formed.
The hoop conjecture for black-hole formation says that,
``black holes with {\it horizons} form when and only when a {\it mass} $m$
gets compacted into a region whose {\it circumference} $C$ in every
direction is $C\leq 4\pi G m$.'' However,
neither the mass, nor notion of the circumference is well-defined.
The most natural definition should be in some sense quasilocal.

In order to understand the physical quantities in these
dynamical processes,
we extend the framework of isolated and
dynamical trapping horizons
to the untrapped regions with the notions of
{\it stationary} and {\it dynamical untrapped hypersurfaces} \cite{Tung2008}.
With these notions, one can give well
motivated definitions of physical quantities such as the energy
and the angular momentum\cite{Tung2008}, and the fluxes \cite{Tung-Yu2008}
of energy and angular momentum
of matter and gravitational radiation falling into the black holes
and other strong gravitating systems.

\section{Stationary and dynamical untrapped hypersurfaces}

We begin with the geometry of an untrapped two-surface $S$ embedded in
a four-dimensional spacetime $M$. Introduce a set of orthonormal vectors
$e_0, e_1, e_2, e_3$ adapted to the two-surface $S$, with $e_0$ and
$e_1$ being the set of timelike and spacelike unit normals to $S$ and
$e_A=(e_2, e_3)$ being tangent to $S$. The extrinsic curvatures of $S$
with respect to $e_0$ and $e_1$ directions are given by
$k(e_0)_{AB}=g(e_B ,\nabla_A e_0)=-\Gamma_{0BA}$ and $k(e_1)_{AB}=
g(e_B, \nabla_A e_1)=-\Gamma_{1BA}$. The connection 1-forms in the
normal bundle are given by $\varpi_{A}=g(e_1, \nabla_A
e_0)=-\Gamma_{01A}$. Here $\Gamma_{IJK}= -g(e_J, \nabla_K e_I)$ are
Ricci rotation coefficients.
The expansion vector $H$, and the dual expansion vector $H_\perp$
are defined with the trace of the extrinsic curvatures $k(e_0)$ and
$k(e_1)$,
\begin{eqnarray} \label{expansionvector}
H &=& k(e_1) e_1 - k(e_0) e_0  ,\\
H_\perp &=& k(e_1) e_0 - k(e_0) e_1  . \label{dualexpansionvector}
\end{eqnarray}
These vectors are
independent of choice of normal frames for the two-surface. They
are uniquely defined given the two-surface $S$ and constitute a set of natural
normal vectors for $S$.
Unlike $H$ and $H_\perp$,
the connection 1-forms in the normal bundle $\varpi_{A}$ depends
on the choice of normal frames for the two-surface. However, for untrapped
surfaces, we can use
the uniquely determined unit normal vectors for the two-surface $S$,
\begin{equation} \label{preferrednormals}
\hat{e}_{0} = \frac{H_\perp}{|H|}, \qquad \hat{e}_{1} = \frac{H}{|H|}
,
\end{equation}
to fix the gauge ($|H|=\sqrt{k(e_1)^2-k(e_0)^2}\neq 0$).
So that $\varpi_A$ is uniquely defined \cite{Szabados1994,
Anco-Tung2002a,
Anco-Tung2002b}.

A two-surface $S$ is {\it trapped}, {\it
untrapped}, or {\it marginal} if
the dual expansion vector
$H_\perp$ is spacelike, timelike, or null respectively,
everywhere on $S$. Note that, on $S$, the trace of the extrinsic curvature is
zero along the direction of the dual expansion vector, i.e.
$k(H_\perp)\vert_S=0$, and we have
\begin{equation} \label{expansionzero}
\pounds_{H_\perp} \varrho \vert_S=0 ,
\end{equation}
where $\varrho$ is the area element of $S$.
This is the key equation for the definition of the
{\it stationary untrapped hypersurfaces}.

\newtheorem{def1}{Definition}

\begin{def1} [stationary untrapped hypersurface]
A smooth timelike hypersurface $\triangle=\mathbb{S}^2\times\mathbf{R}$
is said to be a dynamical untrapped hypersurface if
it can be foliated by a family of closed two-surfaces $S$ such that each
foliation is an untrapped surface.
If on each leaf of the dynamical untrapped hypersurface,
the dual expansion vector $H_\perp$ is tangent to the dynamical untrapped
hypersurface, then it is called a stationary untrapped
hypersurface $\triangle_S$.
\end{def1}
Note that it is the ``hypersurface'' that is ``stationary''.
The actual spacetime can be dynamical and non-stationary.
The key equation (\ref{expansionzero}) for the
stationary untrapped boundary conditions implies the area
of a cross section of $\triangle_S$ is constant along
$H_\perp$.
The definition of stationary untrapped hypersurface keeps
the property of `non-expanding' and generalize the null normal
used in marginal surfaces for isolated
and dynamical trapping horizons, to the dual expansion vector
$H_\perp$ for untrapped surfaces.
Therefore, an alternative name for the
``stationary untrapped hypersurfaces''
might be ``non-expanding untrapped hypersurfaces''.
The dual expansion vector $H_\perp$
plays the role for stationary untrapped
hypersurfaces, which the stationary Killing vector plays for
stationary black holes.
In the limit when the dual expansion vector
$H_\perp$ is null, $S$ reduces to a marginal surface, the
hypersurface reduces to a non-expanding
horizon\cite{Ashtekar-Krishnan-Review}.

\section{Conserved quantities associated with
stationary untrapped hypersurfaces}

The notions of stationary untrapped hypersurfaces
extract the minimal
conditions which are necessary to uniquely define
energy and angular momentum for untrapped strong gravitating systems.
In this section, we shall
derive the conserved quantities by extending
the requirement of the functional differentiability of
the Hamiltonian, considered first by Regge and Teitelboim
\cite{Regge-Teitelboim1974}, for spatial infinity to the finite
spatial two-surfaces.

For a general diffeomorphism-invariant field theory in four
dimensions with a Lagrangian 4-form
${\mathcal L}(\varphi, p)=d\varphi\wedge p-\Lambda(\varphi, p)$, where
$\varphi$ denotes an arbitrary collection of dynamical fields. The
equations of motion are obtained by computing the first variation
of the Lagrangian.
\begin{equation} \label{varL}
\delta {\mathcal L} = d(\delta \varphi \wedge p) + \delta \varphi
\wedge \frac{\delta {\mathcal L}}{\delta \varphi} + \frac{\delta
{\mathcal L}}{\delta p} \wedge \delta p .
\end{equation}
For any diffeomorphism generated by a smooth vector field $\xi$,
we can replace the variational derivative $\delta$ by the Lie
derivative $\pounds_\xi$,
\begin{equation}
\pounds_\xi {\mathcal L} =
d(\pounds_\xi \varphi \wedge p) + \pounds_\xi \varphi
\wedge \frac{\delta {\mathcal L}}{\delta \varphi} + \frac{\delta
{\mathcal L}}{\delta p} \wedge \pounds_\xi p .
\end{equation}
One can then define a conserved Noether current 3-form $J(\xi)$
by
$J(\xi) := \pounds_\xi \varphi \wedge p - i_\xi {\mathcal L}$,
such that the Noether current $d J(\xi)\simeq 0$ is closed
on shell. Locally there exists a 2-form $Q(\xi)=i_\xi \varphi \wedge p$ (the
Noether charge) such that
\begin{equation}
 J(\xi)=\xi^\mu {\mathcal H}_\mu + d Q(\xi).
\end{equation}
On shell, the variation of the Noether current 3-form is given by,
\begin{equation}
\delta J(\xi)= \omega(\varphi,\delta\varphi,
{\rm\pounds}_\xi\varphi) + d i_\xi(\delta \varphi \wedge p),
\end{equation}
where $\omega$ is the presymplectic current 3-form defined by
$
\omega(\varphi, \delta_1\varphi, \delta_2\varphi)=\delta_2 \varphi
\wedge \delta_1 p - \delta_1 \varphi \wedge \delta_2 p.
$
Its integral over a spacelike hypersurface $\Sigma$ defines the presymplectic
form $\Omega$. If
\begin{eqnarray}
\Omega(\varphi, \delta\varphi, \pounds_\xi\varphi) &\equiv&
\int_\Sigma\omega(\varphi, \delta\varphi,\pounds_\xi\varphi)  =\delta
\mathbb{H}(\xi)
\end{eqnarray}
 is a total variation for some function $\mathbb{H}(\xi)$ on the field space, then
$\pounds_\xi
\mathbb{H}(\xi)=0$. $\mathbb{H}(\xi)$ is the Hamiltonian
(conserved quantity) conjugate to $\xi$ \cite{Tung2008,Anco-Tung2002a,Lee-Wald1990,
Nester1991,Wald1993,Chen-Nester-Tung2005}.
One can write the integrand as the exterior derivative of a 2-form.
Therefore the integral is performed over the boundary $S$ of $\Sigma$,
\begin{eqnarray}
\Omega(\varphi, \delta\varphi, \pounds_\xi\varphi) &\equiv&
\oint_{S} \delta Q(\xi) - i_\xi (\delta\varphi \wedge p) =\delta
\mathbb{H}(\xi) .
\end{eqnarray}
The conserved quantity, if it exists, is an integral over this boundary.

For general relativity,
$
 S=\int {\cal L}=\int R^{ab}\wedge
 \ast(\vartheta_a\wedge\vartheta_b) ,
$
where $R^{ab}$ is the
curvature 2-form constructed by the connection 1-form $\Gamma^{ab}$,
 and
 $\vartheta^a$ is the orthonormal
 frame 1-form field.
The presymplectic form is
 given by
\begin{equation}
\Omega(\varphi, \delta\varphi, \pounds_\xi\varphi)=\oint_S
 \frac{1}{2} i_\xi \Gamma^{ab}
\delta(\epsilon_{abcd}\vartheta^c\wedge\vartheta^d )
+\oint_S i_\xi\vartheta^c\wedge\delta\Gamma^{ab}\wedge
\epsilon_{abcd}\vartheta^d .
\end{equation}
Decompose into its normals and
tangents of a two-surface boundary $S$ gives\cite{Tung2008},
\begin{equation} \label{18}
\Omega(\varphi, \delta\varphi, \pounds_\xi\varphi)=
\oint_S 2 \, i_\xi\Gamma^{01} \delta\varrho -
\oint_S 2 \varrho \left(i_\xi \vartheta^0 \delta k(e_1) +i_\xi
\vartheta^1 \delta k(e_0)
 -i_\xi \vartheta^A \delta\varpi{}_A \right)  .
 \end{equation}
Note that the presymplectic form
$\Omega(\varphi, \delta\varphi, \pounds_\xi\varphi)$
depends only on the variation of the area element
(together with the variation of the second and normal fundamental forms).
Not all of
the information of the first fundamental form is required to be fixed.

The stationary untrapped
hypersurface boundary conditions\cite{Tung2008} provides
the ``minimal'' boundary conditions
which are necessary for the presymplectic form
$\Omega(\varphi, \delta\varphi, \pounds_\xi\varphi)$
to be
a total variation for some function $\mathbb{H}(\xi)$,
to define energy and angular momentum.
Quasilocally, we expect these conserved quantities
to depend on the choice of the vector field $\xi$.
The stationary untrapped hypersurface
fixes the boundary conditions for $\xi$ up to
the choice of a quasilocal function of $S$.
In the next three sections we shall
discuss three such choices which give the Hawking energy,
the Brown-York energy and the generalized Hawking energy.

\section{Hawking energy}

For a spherically symmetric stationary untrapped
boundary condition,
we first pick $\xi$ on $S$ to be \cite{Tung2008}
\begin{equation} \label{kodama}
\xi\vert_S=h(\varrho) H_\perp ,
\end{equation}
where $H_\perp$ is the dual expansion vector and $h(\varrho)$ is a
quasilocal function of the area element of the (untrapped or marginal)
two-surface $S$.
Then by equation (\ref{expansionzero}) $\pounds_{\xi} \varrho =0$,
the presymplectic form (\ref{18}) reduced to
\begin{eqnarray}
0=\Omega(\varphi, \pounds_\xi\varphi, \pounds_\xi\varphi) &=&-
\oint_S 2 \varrho h(\varrho) \left( k(e_1) \pounds_{\xi} k(e_1)
- k(e_0) \pounds_{\xi} k(e_0)  \right)  . \nonumber\\
&=&-
\oint_S \varrho h(\varrho) \pounds_{\xi} \left( k(e_1)^2
- k(e_0)^2  \right)  . \nonumber\\
 &=& \pounds_{\xi} \oint_{S}  \left(f(\varrho)-h(\varrho) H^2 \right) \varrho
 \end{eqnarray}
The Hamiltonian conserved quantity
$\mathbb{E}(\xi)$ associated with the vector $\xi$ is given by
\begin{equation}
 \mathbb{E}(\xi)
 =   \oint_S  \, \left(f(\varrho)-h(\varrho) H^2 \right) \varrho .
\end{equation}

The free quasilocal functions of the area element, $f(\varrho)$ and
$h(\varrho)$, can be chosen such that the expression gives ADM mass
at spatial infinity and irreducible mass at marginal surfaces $H=0$.
This can be done by letting $f(\varrho)$ to be $1/(8\pi \mathbb{R})$
and let $h(\varrho)$ to be $\mathbb{R}/(32\pi)$, where $\mathbb{R}$
is the area radius given by
$\mathbb{R}=\sqrt{\frac{1}{4\pi}\oint_S \varrho}$, then
\begin{equation}
 \mathbb{E}_H(\xi)
 = \frac{\mathbb{R}}{2} \left( 1 - \frac{1}{16\pi} \oint_S  \, \,
 H^2 \varrho \right) ,
\end{equation}
which is precisely the Hawking energy \cite{Hawking1968}.
In this case the evolution vector $\xi$ on $S$
\begin{equation}
\xi\vert_S={\mathbb{R}\over 32 \pi} H_\perp
\end{equation}
is precisely the Kodama vector \cite{Kodama1980}.

\section{Brown-York energy}

Alternatively,
we can use the unit dual expansion vector.
For untrapped surfaces ($|H|\neq0$), there is a set of uniquely
determined unit normal vectors for the two-surface given by
equation (\ref{preferrednormals}).
If we choose the evolution vector $\xi$
such that $\xi \vert_S=\hat{e}_0$, this leads to
the energy expression\cite{Tung2008}
\begin{equation}
 \mathbb{E}(\xi)
 =   \oint_S  \, \left(f(\varrho)- k(\hat{e}_1) \right)
 \varrho .
\end{equation}
Here we have only one free quasilocal function $f(\varrho)$.

A natural requirement is
that the expression should give ADM energy at spatial infinity.
By the embedding theorem of Wang and Yau\cite{Wang-Yau2009},
we can embed the two-surface isometrically into Minkowski spacetime,
let  $k_0(e_0)$ and $k_0(e_1)$ be the trace of extrinsic curvatures
with respect to $e_0$ and $e_1$,
for the two-surface in Minkowski spacetime, then the choice
$f(\varrho)=k_0(\hat{e}_1)$ gives the
Brown-York quasilocal energy
\cite{Brown-York1993, Brown-Lau-York2002},
\begin{equation}
 \mathbb{E}_{BY}
  =   \oint_S  \, \left(k_0(\hat{e}_1)- k(\hat{e}_1) \right)
 \varrho .
\end{equation}
Note that
$\hat{e}_{0}$ fails to be defined in the null case.
It seems that this choice is not suitable for the cases
involving dynamical black holes.

\section{Generalized Hawking energy}

For an axisymmetric stationary untrapped boundary condition,
a natural generalization of the evolution vector field $\xi$ on $S$,
is\cite{Tung2008}
\begin{equation} \label{fullvector}
\xi\vert_S=h(\varrho, j) H_\perp - \Omega(\varrho, j)  \psi ,
\end{equation}
which is assumed to be timelike or null, where $h(\varrho, j)$ and
$\Omega(\varrho, j)$ (angular speed) are functions of the area
element $\varrho$ and the angular momentum surface density $j$.
The angular momentum surface density $j$ is 
defined by $j= \psi^A\varpi_A$, where
$\psi $ is a vector tangent to $S$
satisfying $i_\psi \vartheta^0 \vert_S=0$ and $i_\psi \vartheta^1
\vert_S=0$, with the boundary conditions \cite{Ashtekar-Krishnan2003,
Gourgoulhon2005,Booth-Fairhust2005,  Hayward2006,Tung2008,Szabados2006},
\begin{equation} \label{condition1}
\pounds_\psi \varrho \vert_S=0 , \qquad
\pounds_\psi {H_\perp} \vert_S=-\pounds_{H_\perp} \psi \vert_S = 0 .
\end{equation}
The
conserved quantity associated with $\psi$ is
the angular momentum
\begin{equation}
\mathbb{J}(\psi)=\frac{1}{8\pi} \oint_S j \, \varrho=\frac{1}{8\pi}
\oint_S \psi^A \varpi_A \,\varrho .
\end{equation}
The equations (\ref{expansionzero}) and (\ref{condition1}) implies
$\pounds_\xi \varrho \vert_S=0$ and
$\pounds_\xi j \vert_S= 0 $,
the ``stationary untrapped boundary conditions''\cite{Tung2008} are
satisfied. The presympectic form (\ref{18}) then reduced to
\begin{equation}
0=\Omega(\phi,\pounds_\xi\phi, \pounds_\xi\phi)
 = \pounds_{\xi} \oint_{S}  \left(f(\varrho,j)-h(\varrho,j) H^2 \right) \varrho .
 \end{equation}
The Hamiltonian conserved quantity $\mathbb{E}(\xi)$
associated with $\xi$ is given by
\begin{equation}
 \mathbb{E}(\xi)
 =   \oint_S  \, \left(f(\varrho, j)-h(\varrho, j) H^2) \right)
 \varrho .
\end{equation}

By requiring the energy expression gives the horizon energy
for Kerr black holes at marginal surface $H=0$,
we obtain
$f(\varrho,j)={\sqrt{\mathbb{R}^4+4 \mathbb{J}^2}}/{{8\pi
\mathbb{R}^3}} $.
The free quasilocal function $h(\varrho, j)$ can be
chosen such that the
energy is proportional to the Hawking energy and gives ADM mass at
spatial infinity, this implies that
$h(\varrho,j)={\sqrt{\mathbb{R}^4+4 \mathbb{J}^2}}/{32 \pi
\mathbb{R}} $.
Then the Hamiltonian conserved quantity associated with $\xi$,
\begin{equation}
 \mathbb{E}_{GH}
 =\frac{\sqrt{\mathbb{R}^4+4 \mathbb{J}^2}}{2 \mathbb{R}}
 \left( 1 - \frac{1}{16\pi} \oint_S  \, \, H^2
 \varrho \right) ,
 \end{equation}
is the generalized Hawking energy\cite{Tung2008}. The vector on $S$,
\begin{equation}
\xi \vert_S= \frac{\sqrt{\mathbb{R}^4+4 \mathbb{J}^2}}{32 \pi
\mathbb{R}} H_\perp - \Omega \psi,
\end{equation}
is a generalization of the Kodama vector\cite{Tung2008,Kodama1980}.

\section*{Acknowledgments}
This work was partially supported by the
National Natural Science Foundation of China (10771140),
by Shanghai Education Development Foundation (05SG45)
and NCTS, Taiwan.


\end{document}